# Optimum Risk Portfolio and Eigen Portfolio: A Comparative Analysis Using Selected Stocks from the Indian Stock Market

Jaydip Sen and Sidra Mehtab
Praxis Business School, Kolkata, India
email: {jaydip.sen@acm.org, smehtab@acm.org}

**Abstract:** Designing an optimum portfolio that allocates weights to its constituent stocks in a way that achieves the best trade-off between the return and the risk is a challenging research problem. The classical mean-variance theory of portfolio proposed by Markowitz is found to perform sub-optimally on the real-world stock market data since the error in estimation for the expected returns adversely affect the performance of the portfolio. This paper presents three approaches to portfolio design, viz, the minimum risk portfolio, the optimum risk portfolio, and the eigen portfolio, for seven important sectors of the Indian stock market. The daily historical prices of the stocks are scraped from Yahoo Finance website from January 1, 2016, to December 31, 2020. Three portfolios are built for each of the seven sectors chosen for this study, and the portfolios are analyzed on the training data based on several metrics such as annualized return and risk, weights assigned to the constituent stocks, the correlation heatmaps, and the principal components of the eigen portfolios. Finally, the optimum risk portfolios and the eigen portfolios for all sectors are tested on their return over a period of six-month period. The performances of the portfolios are compared and the portfolio yielding the higher return for each sector is identified.

## 1. Introduction

Portfolio Optimization is the task of identifying a set of capital assets and their respective weights of allocation, which optimizes the risk-return pairs. Optimizing a portfolio is a computationally hard problem. The problem gets more complicated if one has to optimize the future return and risk values as a prediction of future stock prices is also an equally challenging task. In fact, there are proponents of the efficient market hypothesis who believe that it is impossible to forecast future stock prices accurately. However, several propositions in the literature show how complex algorithms and predictive models can be effectively used for the precise prediction of future stock prices. Following the seminal work of Markowitz on the minimum-variance portfolio, several propositions have been made for different approaches to portfolio optimization (Markowitz, 1952). A large number of propositions for future stock price prediction also exist that are based on approaches like *multivariate regression*, *autoregressive integrated moving average* (ARIMA), *vector autoregression* (VAR), time series forecasting, machine learning, and deep learning.

This paper presents a systematic approach for building robust portfolios of stocks from seven important sectors of the Indian economy. Seven important sectors of the Indian economy are first chosen, and for each sector, the ten most significant stocks are identified as per their listing in the National Stock Exchange (NSE) of India (NSE, 2021). The historical prices of these 70 stocks are automatically scraped from the Yahoo Finance website using their ticker names. Based on the historical prices over a period of five years, minimum risk portfolios, optimum risk portfolios and eigen portfolios are built for the seven sectors. The performances of the portfolios are evaluated based on their returns after a period of six months. Several other attributes of the portfolios such as the risk, weights assigned to different stocks, the correlation among the constituent stocks, and the principal components in the eigen portfolios are also studied. In addition to providing us with a comparative understanding about the performances of the portfolios, the study also reveals the actual return and volatility values for the sectors, which gives an investor in the stock market an insight into the current return on investment (ROI) and the risk associated with the stocks and their associated sector as a whole.

The work has three major contributions. First, it presents a systematic approach for constructing three portfolios, minimum risk, optimum risk, and eigen portfolios, of stocks for seven important sectors of the Indian economy based on their historical prices over the period of the last five years. These portfolios can serve as illustrative guidance for investments in the stocks of those seven sectors of the Indian stock market. Second, the comparative analysis of the performance of the optimum risk portfolio and the eigen portfolio enables one understand which portfolio design approach works better for what sector. Finally, the results of the study provide an insight into the current profitability and risk associated with investments in the stocks of the seven important sectors of the Indian stock market.

The paper is organized as follows. Section 2 provides a brief discussion on some of the related works. Section 3 provides a description of the data and the methodology followed by us in designing the portfolios and in analyzing the results. Section 4 presents extensive results of the performances of the portfolios and their detailed analysis. Finally, Section 5 concludes the paper and identifies some future research directions.

## 2. Related Work

The literature on stock price prediction models and portfolio optimization is quite rich. Several approaches have been proposed by researchers for accurate prediction of future values of stock prices and using the forecasted results in building robust and optimized portfolios that optimize the returns while minimizing the associated risk. Time series decomposition and econometric approaches like *autoregressive integrated moving average* (ARIMA), Granger causality, *vector autoregression* (VAR) are some of the most popular approaches to future stock price predictions which are used for robust portfolio design (Sen, 2018; Sen & Datta Chaudhuri, 2018; Sen & Datta Chaudhuri, 2017a; Sen & Datta Chaudhuri, 2017b; Sen & Datta Chaudhuri, 2016a; Sen & Data Chaudhuri, 2016b). The use of machine learning, deep learning, and reinforcement learning models for future stock price prediction has been the most popular approach of late (Mehtab & Sen, 2020a; Mehtab & Sen, 2020b; Mehtab et al., 2020a; Mehtab et al., 2020b; Bao et al., 2017; Sen et al., 2021a; Binkowski et al., 2018; Sen & Mehtab, 2021; Mehtab & Sen, 2021). Hybrid models are also proposed that utilize the algorithms and architectures of machine learning and deep learning and exploit the sentiments in the textual sources on the social web (Mehtab & Sen, 2019; Chen et al., 2019; Bollen et al., 2011; Galvez

& Gravano, 2017; Weng et al., 2017). The use of *metaheuristics* algorithms in solving *multi-objective optimization problems* for portfolio management has been proposed in several works (Chen et al., 2018; Macedo et al., 2017; Pai, 2017; Qu et al., 2017). Several modifications of Markowitz's minimum variance portfolio theory have been advocated by imposing a constraint of the purchase limits and on the cardinality (Saborido et al., 2016; Silva et al., 2015; Reveiz-Herault, 2016; Vercher & Bermudez, 2015). The use of fuzzy logic, *genetic algorithms* (GAs), algorithms of swarm intelligence, e.g., *particle swarm optimization* (PSO), are also quite common in portfolio optimization (Garcia et al., 2018; Ertenlice & Kalayci, 2018). The use of *generalized autoregressive conditional heteroscedasticity* (GARCH) in estimating the future volatility of stocks and portfolios is a very popular approach (Sen et al., 2021).

In the present work, we follow three approaches, minimum risk portfolio, optimum risk portfolio, and eigen portfolio, for designing portfolios of stocks of seven important sectors of the Indian stock market. Using the historical prices of the top ten stocks in the NSE for each sector over a period of five years (i.e., from 1 January 2016 to 31 December 2020), the portfolios are built. The actual return values of the optimum risk portfolios are compared with those of the eigen portfolios for each sector over a period of six months. The results are analyzed to identify the portfolio yielding the higher return and also for computing the overall return of each sector.

## 3. Data and Methodology

In this section, we discuss in detail the eight-step approach that we adopted in designing our framework of portfolio design and analysis. proposed system. The eight steps are described in detail in the following sections.

### 3.1. Selection of the sectors and stocks

First, we choose seven important sectors of the Indian economy for our study. The chosen sectors are as follows: *auto*, *banking*, *consumer durable*, *healthcare*, *fast-moving consumer goods* (FMCG), *information technology* (IT), and *metal*. For each sector, we identify the top ten stocks on the basis of their contributions to the sectoral index in the NSE (NSE, 2021). In the NSE, the index value of a given sector is computed based on the index of different stocks listed under the sector. The importance of a stock in a given sector is determined by the corresponding weight of the stock used in the derivation of the overall index of the sector. For each of the seven sectors, the top five stocks are identified based on the weights (in percent) as reported by the NSE (NSE, 2021).

### 3.2. Data acquisition

For each of the seven sectors we have selected sector, we extract the historical prices of the top ten stocks of each sector, using the *YahooFinancials* function in the *yahoofinancial* library of Python. The *ticker* names of the stocks are passed as the parameters to the *YahooFinancials* function with the *start_date* parameter set to '2006-01-01', the *end_date* parameter set to '2020-12-31', and the *time_interval* parameter set to 'daily'. In this way, the stock prices are extracted from the Yahoo Finance website, from 1 January 2016 to 31 December 31 2020 (Yahoo Finance, 2021). The stock data have the following attributes: *open*, *high*, *low*, *close*, *volume*, and *adjusted_close*. Since we focus on univariate analysis in the present work, we select *close* as our variable of interest and don't consider the other variables. We refer to the univariate *close* values of the five stocks for a given sector from 1 January 2016 to 31 December 2020 for training the portfolio models.

### 3.3. Computation of return and volatility

Using the training data for the ten stocks in each sector, we compute the daily return and log return values of each stock of that sector. The daily return values are the percentage changes in the daily close values over successive days, while the *log return* values are the logarithms of the percentage changes in the daily *close* values. For computing the daily return and log return, the *pct_change* function of Python is used. Using the daily return values, we compute the daily volatility and the annual volatility of the ten stocks of each sector. The daily volatility is defined as the standard deviation of the daily return values. The daily volatility, on multiplication by a factor of the *square root* of 250, yields the value of the *annual volatility*. Here, there is a standard assumption of 250 working days in a year for a stock market. The annual volatility value of a stock quantifies the risk associated with stock from the point of view of an investor, as it indicates the amount of variability in its price. For computing the volatility of stocks, the *std* function of Python is used. The daily return values are also aggregated into *annual return* values for each stock for every sector.

### 3.4. Computation of covariance and correlation matrices

After computing the volatilities and return of the stocks, we compute the covariance and the correlation matrices for the ten stocks in each sector using the records in the training dataset. These matrices help us in understanding the strength of association between a pair of stock prices in a given sector. Any pair exhibiting a high value of correlation coefficient indicates a strong association between them. We use the Python functions *cov* and *corr* to compute the covariance and correlation matrices. A good portfolio aims to minimize the risk while optimizing the return. Risk minimization of a portfolio requires identifying stocks that have low correlation among themselves so that a higher diversity can be achieved. Hence, computation and analysis of the covariance and correlation matrices of the stocks are of critical importance.

### 3.5. Computation of the expected return and risk of portfolios

At this step, we proceed towards a deeper analysis of the historical prices of the ten stocks in each of the seven sectors. First, for each sector, we construct a portfolio using the ten stocks, with each stock carrying equal weight. Since there are ten stocks in a sector (i.e., in a portfolio), each stock is assigned a weight of 0.1. Based on the training dataset and using an equal-weight portfolio, we compute the yearly return and risk (i.e., volatility) of each portfolio. The computation of the expected return of a portfolio is done using (1). In (1), $E(R)$ denotes the expected return of a portfolio consisting of *n* stocks, which are denoted as $S_1, S_2, …S_n$. The weights associated with the stocks are represented by $w_i$'s.

$$E(R) = w_1 E(R_{S_1}) + w_2 E(R_{S_2}) + \cdots + w_n E(R_{S_n}) \qquad (1)$$

The yearly return and the yearly volatility of the equal-weight portfolio of each sector are computed using the training dataset. For this purpose, the mean of the yearly return values is derived using the *resample* function in Python with a parameter 'Y'. Yearly volatility values of the stocks in the equal-weight portfolio are derived by multiplying the daily volatility values by the square root of 250, assuming that there are, on average, 250 working days in a year for a stock exchange. The equal-weight portfolio of a sector gives us an idea about the overall profitability and risk associated with each sector over the training period. However, for future investments, their usefulness is very limited. Every stock in a portfolio does not contribute equally to its return and the risk. Hence, we proceed with computations of minimum risk and optimal risk portfolios in the next steps.

### 3.6 Building the minimum risk portfolio

We build the *minimum risk portfolio* for each sector using the records in its training dataset. The minimum risk portfolio is characterized by its *minimum variance*. The variance of a portfolio is a metric computed using the variances of each stock in the portfolio as well as the covariances between each pair of stocks in the portfolio. The variance of a portfolio is computed using (2).

$$Var(P) = \sum_{i=1}^{n} w_i \sigma_i^2 + 2 * \sum_{i,j} w_i * w_j * Cov(i,j) \qquad (2)$$

In (2), $w_i$ and $\sigma_i$ represent the weight associated with stock *i* and the standard deviation of the historical prices of the stock *i*. The covariance among the historical prices of stock *i* and stock *j* is denoted as *Cov(i, j)*. In the present work, there are ten stocks in a portfolio. Hence, 55 terms are involved in the computation of variance of each portfolio, 10 terms for the weighted variances, and 45 terms for the weighted covariances. For building the *minimum risk portfolios*, we need to find the combination of $w_i$'s that minimizes the variance of the portfolio.

For finding the minimum risk portfolio, we first plot the *efficient frontier* for each portfolio. For a given portfolio of stocks, the efficient frontier is the contour with returns plotted along the *y*-axis and the volatility (i.e., risk) on the *x*-axis. The points of an efficient frontier denote the points with the maximum return for a given value of volatility or the minimum value of volatility for a given value of the return. Since, for an efficient frontier, the volatility is plotted along the *x*-axis, the minimum risk portfolio is identified by the leftmost point lying on the efficient frontier. For plotting the contour of the efficient frontier, we randomly assign the weights to the five stocks in a portfolio in a loop and iterate the loop 10,100 times in a Python program. The iteration produces 10,000 points, each point representing a portfolio. The minimum risk portfolio is identified by detecting the leftmost point on the efficient frontier.

### 3.7 Computing the optimum risk portfolio

The investors in the stock markets are usually not interested in the minimum risk portfolios as the return values are usually low. In most cases, the investors are ready to incur some amount of risk if the associated return values are high. To compute the optimum risk portfolio, we use the metric *Sharpe Ratio* of a portfolio. The Sharpe Ratio of a portfolio is given by (3).

$$Sharpe\ Ratio = \frac{R_c - R_f}{\sigma_c} \qquad (3)$$

In (3), $R_c$, $R_f$, and $\sigma_c$ denote the return of the current portfolio, the risk-free portfolio, and the standard deviation of the current portfolio, respectively. Here, the risk-free portfolio is a portfolio with a volatility value of 1%. The *optimum-risk portfolio* is the one that maximizes the *Sharpe Ratio* for a set of stocks. This portfolio makes an optimization between the return and the risk of a portfolio. It yields a substantially higher return than the minimum risk portfolio, with a very nominal increase in the risk, and hence, maximizing the value of the Sharpe ratio. We identify the optimal portfolio using the *idmax* function in Python over the set of the Sharpe Ratio values computed for all the points of an efficient frontier.

### 3.8 Computing the eigen portfolio

The design of eigen portfolios of stocks is based on the concept of *principal component analysis* (PCA). PCA is a very popular unsupervised learning method for reducing the dimensionality of a dataset while retaining the inherent variance in the data. PCA computes a new set of variables, called *components*, which are linear combinations of the original variables in the data. Orthogonality is an important property of the components. In other words, the correlation between a pair of components is always zero, even when the original variables may have a substantial correlation among them. Usually, the number of components is much lower than the number of original variables so that a substantial reduction in dimensionality is achieved. The number of components is a *hyperparameter* of a PCA algorithm. In a nutshell, PCA identifies the directions of maximum variance in data with high dimensionality and projects it onto a new space with equal or fewer dimensions than the original space. In Figure 1, the scatter plot of the two original data points is shown in red color. The two orthogonal components explaining the variance in the data are shown in black color. The lengths of the components are shown in proportion to their power of explanation of the variance.

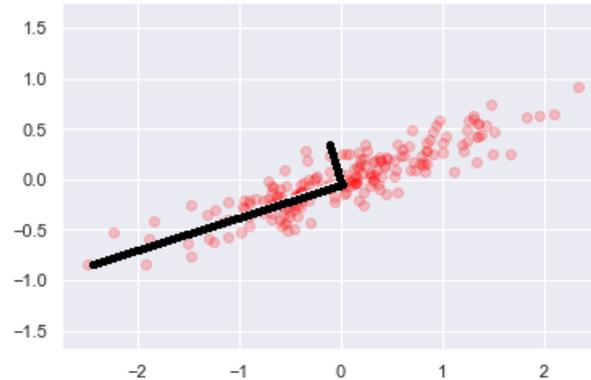

**Fig 1.** Two principal components in two-dimensional data. The black lines exhibit the two principal components, and the red dots represent the data points.

Principal components in a dataset can be found using three approaches: (i) *eigen decomposition*, (ii) *singular value decomposition* (SVD), and (iii) *kernel principal component analysis* (KPCA). The construction of the eigen portfolio is based on the principles of eigen decomposition.

The eigen decomposition approach consists of four steps. They are as follows. First, a covariance matrix for the original variables (after they are standardized) in the data is computed. Second, the eigenvectors of the covariance matrix are obtained. The eigenvectors exhibit the directions of maximum variance in the data. Third, the eigenvalues are computed. The eigenvalues provide us with the magnitudes of the principal components. In the case of a dataset containing *n* variables, the dimension of the covariance matrix will be *n* x *n*. The covariance matrix will yield an eigenvector of *n* values and *n* eigenvalues. Finally, the eigenvalues are sorted in decreasing order to rank the corresponding eigenvectors, and *k* eigenvectors are selected corresponding to the *k* largest eigenvalues. The data are now transformed from the original *d* dimension to the new *k* dimension where $k \leq d$.

The *sklearn* library of Python has an efficient implementation of PCA, which has been used for designing the eigen portfolios of the ten stocks of each of the seven sectors. The desired number of principal components is used as the hyperparameter to the PCA function defined in the decomposition module of *sklearn*. The following steps are performed in constructing the eigen portfolios for the seven sectors.

First, the correlation matrix for the stocks in each sector is computed. For better visualization, the correlation matrices are presented in a heatmap mode using the *heatmap* function defined in the *seaborn* library. The correlation heatmap provides us with a very effective visualization of the correlation among the stocks in a sector.

In the second step, data cleaning and data transformation are carried out. For data cleaning, we check for the existence of any NA values in the rows. Any stock (i.e., any column) having more than 25% missing values is dropped. Other missing values are imputed using the mean imputation technique, i.e., the missing values are substituted by the mean value of the whole column. The dataset values are standardized using the *StandardScaler* function of the *sklearn* library. Standardization of data values with a zero mean and unit standard deviation is an important requirement for the effective performance of a dimensionality reduction algorithm.

In the third step, we use the PCA function of the *sklearn* library to generate the principal components. The number of components is specified as a hyperparameter to the PCA function. The variances explained by the components and their cumulative explanations are noted. The explained variance by the first component will be the highest, and the percentage of variance explained will decrease consistently for the subsequent components. The eigenvectors with the lowest eigenvalues explain the least amount of variance in the data, and therefore, those eigenvalues are dropped.

In the fourth step, we look into the weights of the factors on each principal component. To do so, we construct five portfolios and assign the weights to the stocks as the first five principal components.

In the final step, we select the best eigen portfolio. The best eigen portfolio is the one that yields the highest value of the Sharpe ratio. As already explained earlier, a high Sharpe ratio implies a higher return with a lower risk. The annualized Sharpe ratio value is computed as the ratio of the yearly return to the annual volatility. For identifying the portfolio with the highest Sharpe ratio, we use a Python function that iterates over a lop and computes the principal component weights for each eigen portfolio and computes is Sharpe ratio. After the eigen portfolio with the highest Sharpe ratio is identified, another Python function is used to check the weights assigned by the portfolio to its constituent stocks.

## 4. Experimental Results

In this section, we present the results on the seven portfolios and analyze the results in detail. As mentioned earlier, the chosen seven sectors are (i) *auto*, (ii) *banking*, (iii) *consumer durable*, (iv) *healthcare*, (v) *fast-moving consumer goods* (FMCG), (vi) *information technology* (IT), and (vii) *metal*. In implementing the portfolio models, we use Python 3.7.4 language and its libraries. The models are trained and validated on the Google Colab GPU runtime environment (Google Colab, 2021).

### 4.1 Auto sector stocks

The top ten NSE-listed stocks of the auto sector and their respective weights in percent values in the computation of the overall index of the sector as per the report released by NSE on 30 June 2021 are mentioned as follows (NSE, 2021). The company names are followed by their contributions in the overall index of the sector in percent figures are mentioned: (1) Maruti Suzuki India: 20.41, (2) Mahindra & Mahindra: 15.21, (3) Tata Motors: 12.44, (4) Bajaj Auto: 11.00, (5) Hero MotoCorp: 7.70, (6) Eicher Motors: 7.61, (7) Bharat Forge: 3.91, (8) Balkrishna Industries: 3.17, (9) Ashok Leyland: 3.60, and (10) MRF: 3.40.

Table 1 depicts the annual return and the annual risk (i.e., volatility) for the stocks of the auto sector over the training period, i.e., from 1 January 2016 to values 31 December 2020. Tata Motors is found to exhibit the lowest annual return and the highest annualized risk. While Maruti Suzuki yields the highest annual return, the lowest risk is found for Bajaj Auto.

**Table 1.** The return and the risk of the auto sector stocks

| Stocks | Annual Return (%) | Annual Risk (%) |
|---|---|---|
| Maruti Suzuki | 15.41 | 31.36 |
| Mahindra & Mahindra | 8.88 | 31.79 |
| Tata Motors | -15.46 | 47.34 |
| Bajaj Auto | 8.40 | 26.10 |
| Hero MotoCorp | 3.00 | 30.38 |
| Eicher Motors | 6.13 | 34.31 |
| Bharat Forge | 8.26 | 37.52 |
| Balakrishna Industries | 4.11 | 36.80 |
| Ashok Leyland | 7.65 | 44.54 |
| MRF | 13.88 | 27.45 |

**Table 2.** The portfolios of the auto sector stocks

| Stocks | Min Risk Portfolio | Opt. Risk Portfolio | Eigen Portfolio |
|---|---|---|---|
| Maruti Suzuki | 0.0118 | 0.0358 | 0.1200 |
| Mahindra & Mahindra | 0.0953 | 0.0607 | 0.1000 |
| Tata Motors | 0.0131 | 0.0099 | 0.1100 |
| Bajaj Auto | 0.2403 | 0.0530 | 0.1000 |
| Hero MotoCorp | 0.1562 | 0.0797 | 0.1100 |
| Eicher Motors | 0.0989 | 0.1337 | 0.1000 |
| Bharat Forge | 0.0256 | 0.0032 | 0.1000 |
| Balakrishna Industries | 0.0583 | 0.3235 | 0.0700 |
| Ashok Leyland | 0.0003 | 0.0650 | 0.1000 |
| MRF | 0.3002 | 0.2355 | 0.0900 |

Table 2 presents the allocation of weights to different stocks of the auto sector using three portfolio design approaches – (i) minimum risk portfolio, (ii) optimum risk portfolio, and (iii) eigen risk portfolio. The sum of weights for each of the three cases is 1. The stock which receives the highest allocation of weight as per the minimum risk, optimum risk, and the eigen portfolio are MRF, Balakrishna Industries, and Maruti Suzuki, respectively.

Table 3 shows the risk and the return values associated with the three portfolios of the auto sector stocks. These values are computed using the prices of the stocks over the training period. It is found that among the three portfolios, the eigen portfolio approach indicated the lowest return and the lowest risk, while the optimum risk portfolio yielded the highest values for the return and risk. Note that even when we have computed the return and the risk of the minimum risk portfolio and the respective weights for allocation to different stocks under this portfolio design, we will not test the performance of this portfolio. This is because of the fact that the minimum risk portfolios usually have a very low return, and hence, they are not followed by the investors in the stock market.

**Table 3.** The return and the risk values of the auto sector portfolios

| Metric | Min Risk Portfolio | Opt. Risk Portfolio | Eigen Portfolio |
| --- | --- | --- | --- |
| Portfolio Return | 10.69% | 19.52% | 8.82% |
| Portfolio Risk | 21.31% | 23.50% | 16.50% |

**Table 4.** The actual return of the optimum portfolio of the auto sector

| Stock | Date: January 1, 2021 | | | Date: July 1, 2021 | | Return |
| --- | --- | --- | --- | --- | --- | --- |
| | Price/Stock | Amount Invested | No. of Stocks | Price/Stock | Actual Values of Stocks | |
| Maruti Suzuki | 7691 | 3580 | 0.47 | 7584 | 3530 | |
| Mahindra | 732 | 6070 | 8.29 | 779 | 6460 | |
| Tata Motors | 187 | 990 | 5.29 | 344 | 1821 | |
| Bajaj-Auto | 3481 | 5300 | 1.52 | 4205 | 6402 | |
| Hero MotoCorp | 3103 | 7970 | 2.57 | 2923 | 7508 | 18.17% |
| Eicher Motors | 2543 | 13370 | 5.26 | 2675 | 14064 | |
| Bharat Forge | 538 | 320 | 0.59 | 766 | 456 | |
| Balkrishna Ind. | 1642 | 32350 | 19.70 | 2275 | 44821 | |
| Ashok Leyland | 99 | 6500 | 65.66 | 122 | 8010 | |
| MRF | 76022 | 23550 | 0.31 | 81018 | 25098 | |
| Total | | 100000 | | | 118170 | |

Table 4 shows the return for an investor who followed the optimum risk portfolio approach and invested a total amount of INR 100000 on 1 January 2021. Note that the total amount here is just an example. The overall return percent will not be affected by the amount invested. As per Table 4, an investor who followed the optimum risk portfolio approach would invest the respective amount in the stocks based on the proportion indicated by the portfolio design. These values are noted under the column "Amount Invested" on 1 January 2021. The market values of the stocks are noted in the column "Price/Stock" on 1 January 2021. Using these values, the number of shares purchased by the investor for each stock are computed and are listed under the column "No. of Stocks". After six months, the actual price of each stock is noted and listed in the column "Price/Stock" on 1 July 2021. For a given stock, its actual rice on 1 July 2021 is multiplied with the corresponding no. of shares to compute the actual value of the stocks. The actual values of all ten stocks are summed up to find the total value of the stocks on 1 July 2021. The return for the six months period under the optimum risk portfolio is found to be 18.17%. Figure 2 exhibits the efficient frontier of the portfolios of the auto sector stocks.

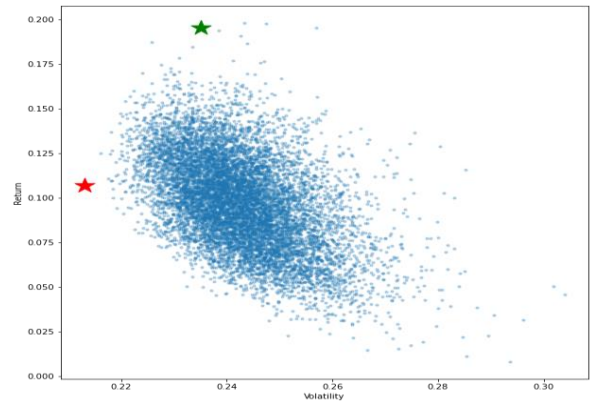

**Fig. 2.** The minimum risk portfolio (the red star) and the optimum risk portfolio (the green star) for the auto sector on historical stock prices from 1 January 2016 to 31 December 2020 (The risk is plotted along the *x*-axis and the return along the *y*-axis)

**Table 5.** The actual return of the eigen portfolio of the auto sector

| Stock | Date: January 1, 2021 | | | Date: July 1, 2021 | | Return |
|---|---|---|---|---|---|---|
| | **Price/Stock** | Amount Invested | No. of Stocks | Price/Stock | Actual Values of Stocks | |
| Maruti Suzuki | 7691 | 12000 | 1.56 | 7584 | 11833 | |
| Mahindra | 732 | 10000 | 13.66 | 779 | 10642 | |
| Tata Motors | 187 | 11000 | 58.82 | 344 | 20235 | |
| Bajaj-Auto | 3481 | 10000 | 2.87 | 4205 | 12080 | |
| Hero MotoCorp | 3103 | 11000 | 3.54 | 2923 | 10362 | |
| Eicher Motors | 2543 | 10000 | 3.93 | 2675 | 10519 | 21.52% |
| Bharat Forge | 538 | 10000 | 18.59 | 766 | 14238 | |
| Balkrishna Ind. | 1642 | 7000 | 4.26 | 2275 | 9699 | |
| Ashok Leyland | 99 | 10000 | 101.01 | 122 | 12323 | |
| MRF | 76022 | 9000 | 0.12 | 81018 | 9591 | |
| Total | | 100000 | | | 121522 | |

Table 5 shows the performance of the eigen portfolio of the auto sector stocks. In designing the eigen portfolio, seven components were chosen. The choice of the number of components was made based on the cumulative explanation of the variance of the close values of the stocks. The threshold minimum cumulative percentage is chosen to be 80%. A minimum number of seven components was found to be necessary for explaining 83.93% of the variance of the close values of the stocks. Figure 3 depicts the variance explained by the individual components and their cumulative explained variance. Figure 4 shows the correlation heatmap of the auto sector stocks, and Figure 5 depicts the weights assigned to the stocks of the auto sector by the five eigen portfolios based on the first five principal components, as discussed in Section 3.8. Investing a total amount of INR 100000 on 1 January 2021 as recommended by the weights in Table 2, an investor buys the number of shares of different stocks mentioned under the column "No of Stocks". Based on the actual values of the stocks on 1 July 2021, the total values of the stocks are computed. Finally, the overall return in percent for the eigen portfolio is computed. The overall return for the eigen portfolio for the auto sector stocks is 21.52%, which is higher than the corresponding figure for the optimum risk portfolio.

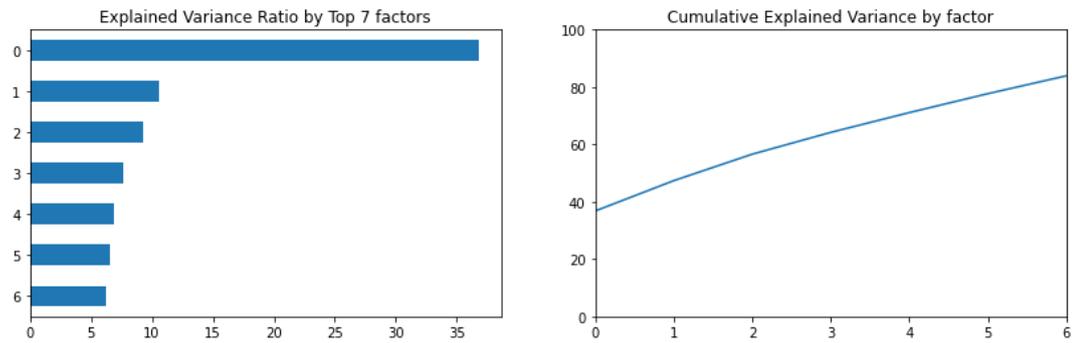

**Fig. 3.** The percentage of variance explained by seven components in the eigen portfolio and the cumulative explained variance by the same components of the eigen portfolio of the auto sector based on the historical prices of the stocks from 1 January 2016 to 31 December 2020.

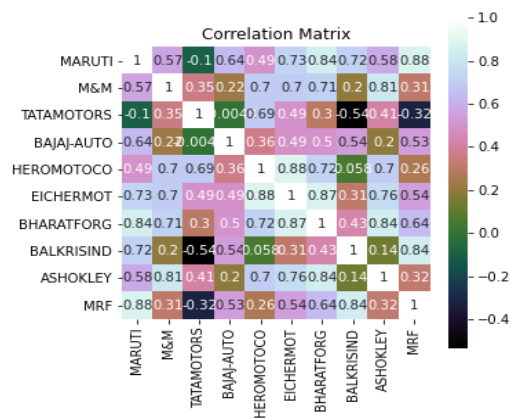

**Fig. 4.** The correlation heatmap for the stocks of the auto sector

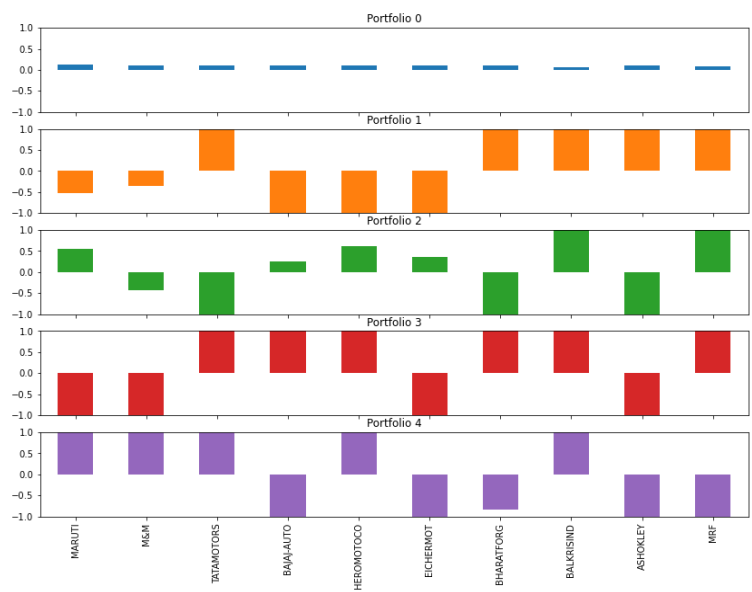

**Fig. 5.** Five eigen portfolios of the auto sector stocks with weights of each stock based on the first five principal components

## 4.2 Banking sector stocks

The top ten NSE-listed stocks of the banking sector and their respective weights in percent values in the computation of the overall index of the sector as per the report released by NSE on 30 June 2021 are as follows (NSE, 2021). (1) HDFC Bank: 28.78, (2) ICICI Bank: 21.06, (3) Kotak Mahindra Bank: 12.07, (4) Axis Bank: 12.06, (5) State Bank of India: 12.00, (6) IndusInd Bank: 5.36, (7) AU Small Finance Bank: 1.98, (8) Bandhan Bank: 1.95, (9) Federal Bank: 1.52, and (10) IDFC First Bank: 1.26.

Table 6 shows the annual return and the annual risk for the banking sector stocks over the training period. The lowest return and the highest risk are found for the Bandhan Bank stock. Kotak Mahindra Bank is found to have the highest annual return, while the risk for the HDFC Bank is found to be the lowest.

**Table 6.** The return and the risk of the banking sector stocks

| Stocks | Annual Return (%) | Annual Risk (%) |
|---|---|---|
| HDFC Bank | 25.27 | 23.61 |
| ICICI Bank | 24.43 | 36.66 |
| Kotak Mahindra Bank | 29.65 | 28.30 |
| Axis Bank | 9.95 | 38.86 |
| State Bank of India | 3.78 | 37.79 |
| IndusInd Bank | 0.06 | 46.37 |
| AU Small Finance Bank | 9.51 | 42.28 |
| Bandhan Bank | -13.86 | 61.56 |
| Federal Bank | 4.76 | 40.81 |
| IDFC First Bank | -11.11 | 40.39 |

Table 7 shows the weights allocated to different stocks of the banking sector using three portfolio design approaches. The stock with the highest allocation of weight as per both minimum risk and optimum risk portfolios is HDFC Bank. For the eigen portfolio, the stocks which received the highest weights are ICICI Bank, Axis Bank, and State Bank of India.

**Table 7.** The portfolios of the banking sector stocks

| Stocks | Min Risk Portfolio | Opt. Risk Portfolio | Eigen Portfolio |
|---|---|---|---|
| HDFC Bank | 0.3805 | 0.2674 | 0.09 |
| ICICI Bank | 0.0487 | 0.2177 | 0.12 |
| Kotak Mahindra Bank | 0.2373 | 0.2240 | 0.08 |
| Axis Bank | 0.0386 | 0.0042 | 0.12 |
| State Bank of India | 0.0340 | 0.0228 | 0.12 |
| IndusInd Bank | 0.0981 | 0.0807 | 0.10 |
| AU Small Finance Bank | 0.0362 | 0.0512 | 0.06 |
| Bandhan Bank | 0.0025 | 0.0338 | 0.09 |
| Federal Bank | 0.0006 | 0.0298 | 0.11 |
| IDFC First Bank | 0.1237 | 0.0684 | 0.11 |

The risk and the return values associated with the three portfolios of the stocks of the banking sector are exhibited in Table 8. As discussed earlier, these values are computed using the historical prices of the stocks over the training period. The eigen portfolio for the banking sector stocks is found to have yielded the highest values for both return and risk.

**Table 8.** The return and the risk values of the banking sector portfolios

| Metric | Min Risk Portfolio | Opt. Risk Portfolio | Eigen Portfolio |
|---|---|---|---|
| Portfolio Return | 17.29 | 18.24 | 42.37% |
| Portfolio Risk | 23.63 | 25.00 | 55.36% |

**Table 9.** The actual return of the optimum portfolio of the banking sector

| Stock | Date: January 1, 2021 | | | Date: July 1, 2021 | | Return |
|---|---|---|---|---|---|---|
| | Price/Stock | Amount Invested | No. of Stocks | Price/Stock | Actual Value of Stocks | |
| HDFC Bank | 1425 | 26740 | 18.76 | 1487 | 27903 | |
| ICICI Bank | 528 | 21770 | 41.23 | 631 | 26017 | |
| Kotak Mahindra Bank | 1994 | 22400 | 11.23 | 1716 | 19277 | |
| Axis Bank | 624 | 420 | 0.67 | 746 | 502 | 8.66% |
| State Bank of India | 279 | 2280 | 8.17 | 420 | 3432 | |
| IndusInd Bank | 900 | 8070 | 8.97 | 1008 | 9038 | |
| AU Small Fin Bank | 875 | 5120 | 5.85 | 1023 | 5986 | |
| Bandhan Bank | 400 | 3380 | 8.45 | 326 | 2755 | |
| Federal Bank | 68 | 2980 | 43.82 | 86 | 3769 | |
| IDFC First Bank | 37 | 6840 | 184.86 | 54 | 9983 | |
| Total | | 100000 | | | 108662 | |

**Table 10.** The actual return of the eigen portfolio of the banking sector

| Stock | Date: January 1, 2021 | | | Date: July 1, 2021 | | Return |
|---|---|---|---|---|---|---|
| | Price/Stock | Amount Invested | No. of Stocks | Price/Stock | Actual Value of Stocks | |
| HDFC Bank | 1425 | 9000 | 6.32 | 1487 | 9392 | |
| ICICI Bank | 528 | 12000 | 22.73 | 631 | 14341 | |
| Kotak Mahindra Bank | 1994 | 8000 | 4.01 | 1716 | 6885 | |
| Axis Bank | 624 | 12000 | 19.23 | 746 | 14346 | |
| State Bank of India | 279 | 12000 | 43.01 | 420 | 18065 | 18.54% |
| IndusInd Bank | 900 | 10000 | 11.11 | 1008 | 11200 | |
| AU Small Fin Bank | 875 | 6000 | 6.86 | 1023 | 7015 | |
| Bandhan Bank | 400 | 9000 | 22.50 | 326 | 7335 | |
| Federal Bank | 68 | 11000 | 161.76 | 86 | 13912 | |
| IDFC First Bank | 37 | 11000 | 297.30 | 54 | 16054 | |
| Total | | 100000 | | | 118543 | |

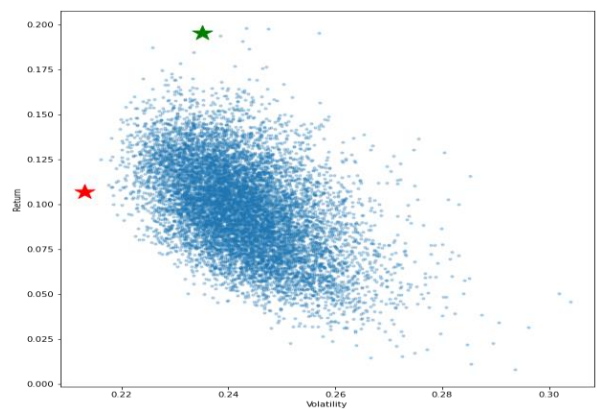

**Fig. 6.** The minimum risk portfolio (the red star) and the optimum risk portfolio (the green star) for the banking sector on historical stock prices from 1 January 2016 to 31 December 2020 (The risk is plotted along the *x*-axis and the return along the *y*-axis)

Table 9 shows the return for an investor who followed the optimum risk portfolio approach and invested a total amount of INR 100000 on 1 January 2021. As in the auto sector portfolios, the number of shares for each stock purchased on 1 January 2021 is computed based on the actual price of the stock and the amount invested. Similarly, the total value of the stocks on 1 July 2021 is derived, and the overall return is computed. For the six-month period, i.e., from 1 January 2021 to 1 July 2021, the return for the banking sector optimum risk portfolio is found to be 8.66%. The efficient portfolio of the banking sector stocks is shown in Figure 6.

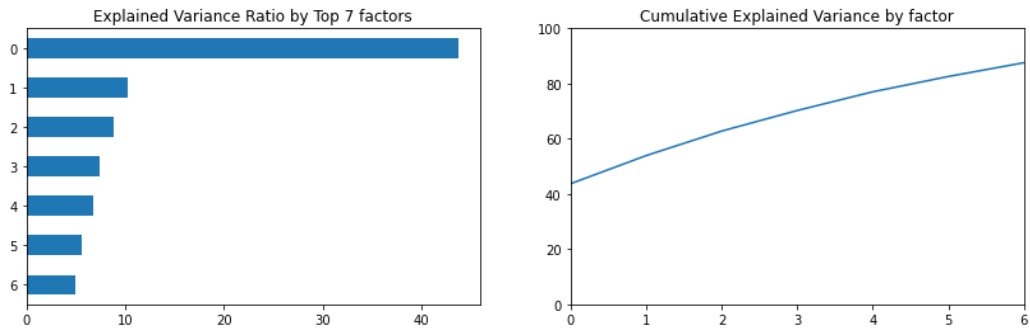

**Fig. 7.** The percentage of variance explained by seven components in the eigen portfolio and the cumulative explained variance by the same components of the eigen portfolio of the banking sector based on the historical prices of the stocks from 1 January 2016 to 31 December 2020.

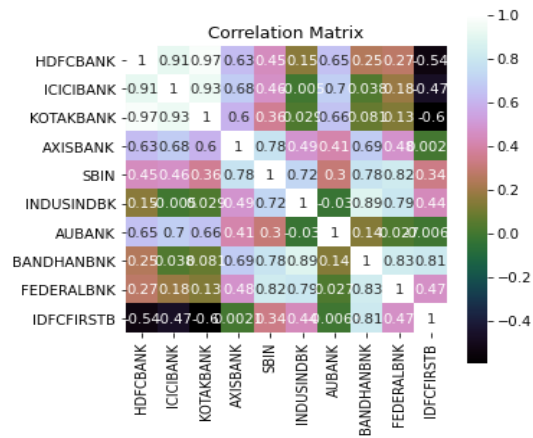

**Fig. 8.** The correlation heatmap for the stocks of the banking sector

Table 10 depicts the results of the eigen portfolio for the stocks of the banking sector. As in the case of the auto sector, seven components are used in designing the eigen portfolio. The cumulative explanation of the variance by the seven components is 87.55%. Figure 7 exhibits the variance explained by the components and their cumulative explanation. Figure 8 shows the correlation heatmap of the banking sector stocks, and Figure 9 depicts the weights assigned to the stocks of the banking sector by the five eigen portfolios based on the first five principal components, as discussed in Section 3.8. The six-month return yielded by the eigen portfolio for the banking sector stocks is 18.54%, which is substantially higher than the return from the optimum risk portfolio.

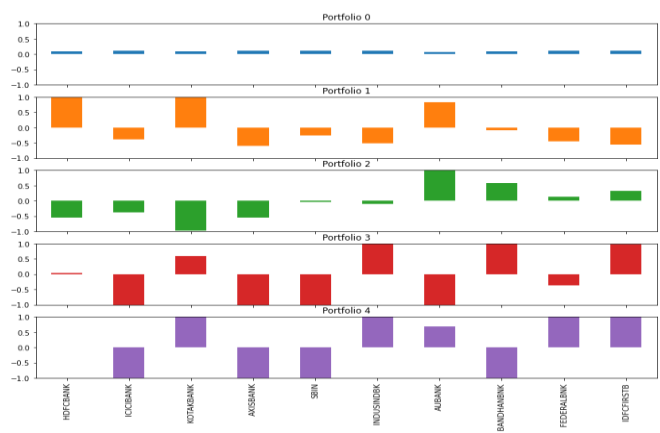

**Fig. 9.** Five eigen portfolios of the banking sector stocks with weights of each stock based on the first five principal components

### 4.3 Consumer Durable sector stocks

The top ten NSE-listed stocks of the consumer durable sector and their respective weights in percent values in the computation of the overall index of the sector as per the report released by NSE on 30 June 2021 are as follows (NSE, 2021). (1) Titan Company: 32.22, (2) Crompton Greaves Consumer Electricals: 11.43, (3) Havells India: 10.95, (4) Voltas: 10.56, (5) Dixon Technologies India: 7.04, (6) Bata India: 4.30, (7) Relaxo Footwears: 3.70, (8) Kajaria Ceramics: 3.61, (9) Rajesh Exports: 3.42, and (10) Whirlpool India: 3.16.

**Table 11.** The return and the risk of the consumer durable sector stocks

| Stocks | Annual Return (%) | Annual Risk (%) |
|---|---|---|
| Titan Company | 57.34 | 33.71 |
| Crompton Greaves | 32.59 | 32.94 |
| Havells India | 30.16 | 30.98 |
| Voltas | 31.92 | 32.65 |
| Dixon Technologies | 95.09 | 40.75 |
| Bata India | 40.60 | 29.48 |
| Relaxo Footwears | 43.73 | 30.83 |
| Kajaria Ceramics | 15.52 | 32.79 |
| Rajesh Exports | 8.82 | 28.01 |
| Whirlpool of India | 36.53 | 32.49 |

**Table 12.** The portfolios of the consumer durable sector stocks

| Stocks | Min Risk Portfolio | Opt Risk Portfolio | Eigen Portfolio |
|---|---|---|---|
| Titan Company | 0.0844 | 0.1712 | 0.12 |
| Crompton Greaves | 0.1387 | 0.0714 | 0.08 |
| Havells India | 0.0899 | 0.0025 | 0.13 |
| Voltas | 0.0044 | 0.0361 | 0.13 |
| Dixon Technologies | 0.0233 | 0.2234 | 0.08 |
| Bata India | 0.0703 | 0.0491 | 0.12 |
| Relaxo Footwears | 0.1394 | 0.2139 | 0.08 |
| Kajaria Ceramics | 0.1323 | 0.0468 | 0.11 |
| Rajesh Exports | 0.1990 | 0.0874 | 0.05 |
| Whirlpool of India | 0.1182 | 0.0982 | 0.10 |

Table 11 exhibits the annual return and the annual risk for the consumer durable sector stocks for the period 1 January 2016 to 31 December 2020. The highest return and the highest risk are found to be yielded by Dixon Technologies. On the other hand, Rajesh Exports exhibits the lowest return and the lowest risk over the same period.

Table 12 depicts the weights allocated to different stocks of the consumer durable sector by the three portfolio design approaches. The stock with the highest allocation of weight as per the minimum risk and the optimum risk portfolios are Rajesh Exports and Dixon Technologies, respectively. For the eigen portfolio, the stocks which received the highest allocation of weights are Havells India and Voltas.

The risk and the return values for three portfolios for the consumer durable sector stocks based on the historical prices of the stocks from 1 January 2016 to 31 December 2021 are presented in Table 13. The eigen portfolio for the consumer durable sector is found to have yielded the highest values for both return and risk.

**Table 13.** The return and the risk values of the consumer durable sector portfolios

| Metric | Min Risk Portfolio | Opt. Risk Portfolio | Eigen Portfolio |
|---|---|---|---|
| Portfolio Return | 31.51 | 51.05 | 72.32 |
| Portfolio Risk | 17.56 | 19.91 | 40.12 |

**Table 14.** The actual return of the optimum portfolio of the consumer durable sector

| Stock | Date: January 1, 2021 | | | Date: July 1, 2021 | | Return |
|---|---|---|---|---|---|---|
| | Price/Stock | Amount Invested | No. of Stocks | Price/Stock | Actual Value of Stocks | |
| Titan Company | 1559 | 17120 | 10.98 | 1740 | 19108 | |
| Crompton Greaves | 378 | 7140 | 18.89 | 429 | 8103 | |
| Havells India | 910 | 250 | 0.27 | 989 | 272 | |
| Voltas | 831 | 3610 | 4.34 | 1013 | 4401 | |
| Dixon Technologies | 2724 | 22340 | 8.20 | 4412 | 36184 | 27.77% |
| Bata India | 1574 | 4910 | 3.12 | 1597 | 4982 | |
| Relaxo Footwears | 828 | 21390 | 25.83 | 1139 | 29424 | |
| Kajaria Ceramics | 709 | 4680 | 6.60 | 988 | 6522 | |
| Rajesh Exports | 485 | 8740 | 18.02 | 565 | 10182 | |
| Whirlpool of India | 2615 | 9820 | 3.76 | 2287 | 8588 | |
| Total | | 100000 | | | 127766 | |

**Table 15.** The actual return of the eigen portfolio of the consumer durable sector

| Stock | Date: January 1, 2021 | | | Date: July 1, 2021 | | Return |
|---|---|---|---|---|---|---|
| | Price/Stock | Amount Invested | No. of Stocks | Price/Stock | Actual Value of Stocks | |
| Titan Company | 1559 | 12000 | 7.70 | 1740 | 13393 | |
| Crompton Greaves | 378 | 8000 | 21.16 | 429 | 9079 | |
| Havells India | 910 | 13000 | 14.29 | 989 | 14129 | |
| Voltas | 831 | 13000 | 15.64 | 1013 | 15847 | |
| Dixon Technologies | 2724 | 8000 | 2.94 | 4412 | 12957 | 18.49% |
| Bata India | 1574 | 12000 | 7.62 | 1597 | 12175 | |
| Relaxo Footwears | 828 | 8000 | 9.66 | 1139 | 11005 | |
| Kajaria Ceramics | 709 | 11000 | 15.51 | 988 | 15329 | |
| Rajesh Exports | 485 | 5000 | 10.31 | 565 | 5825 | |
| Whirlpool of India | 2615 | 10000 | 3.82 | 2287 | 8746 | |
| Total | | 100000 | | | 118485 | |

Table 14 depicts the return for the optimum risk portfolio approach. As it was done for the two sectors discussed previously, the number of shares for each stock purchased on 1 January 2021 is determined on the basis of the actual share price of the stock and the amount invested. The total value of the stocks on 1 July 2021 is calculated, and the overall return is derived. For the six-month period, i.e., from 1 January 2021 to 1 July 2021, the return for the consumer durable sector optimum risk portfolio is found to be 27.77%. The efficient portfolio of the consumer durable sector is shown in Figure 10.

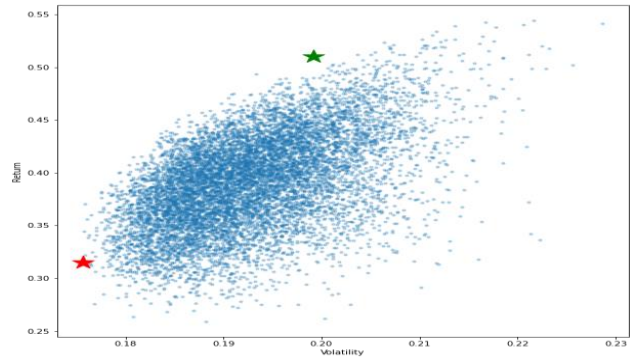

**Fig. 10.** The minimum risk portfolio (the red star) and the optimum risk portfolio (the green star) for the consumer durable sector on historical stock prices from 1 January 2016 to 31 December 2020 (The risk is plotted along the *x*-axis and the return along the *y*-axis)

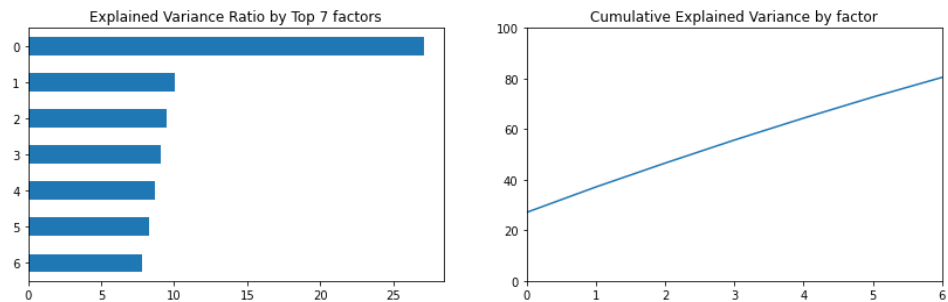

**Fig. 11.** The percentage of variance explained by seven components in the eigen portfolio and the cumulative explained variance by the same components of the eigen portfolio of the consumer durable sector based on the historical prices of the stocks from 1 January 2016 to 31 December 2020.

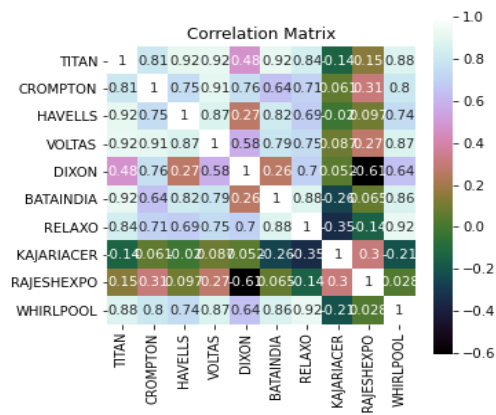

**Fig 12.** The correlation heatmap for the stocks of the consumer durable sector

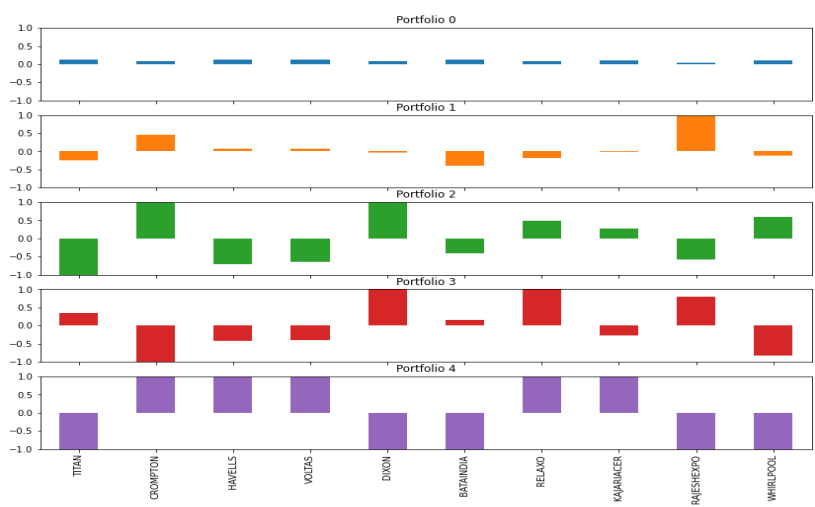

**Fig. 13.** Five eigen portfolios of the consumer durable sector stocks with weights of each stock based on the first five principal components

The results of the eigen portfolio for the consumer durable sector stocks are presented in Table 15. Seven components are used in building the eigen portfolio. The cumulative explanation of the variance by the seven components is 80.47%. Figure 11 depicts the variance and cumulative variance explained by the seven components. Figure 12 shows the correlation heatmap of the consumer durable sector stocks, and Figure 13 depicts the weights assigned to the stocks of the consumer durable sector by the five eigen portfolios, as discussed in Section 3.8. The six-month return for the consumer durable sector as yielded by the eigen portfolio is 18.49%, which is lower than the return produced by the optimum risk portfolio for this sector.

### 4.4 FMCG sector stock

The top ten NSE-listed stocks of the FMCG sector and their respective weights in percent values in the computation of the overall index of the sector as per the report released by NSE on 30 June 2021 are as follows (NSE, 2021). (1) Hindustan Unilever: 29.29, (2) ITC: 23.52, (3) Nestle India: 8.35, (4) Tata Consumer Products: 6.00, (5) Britannia Industries: 5.72, (6) Dabur India: 4.40, (7) Godrej Consumer Products: 4.37, (8) Marico: 3.64, (9) Jubilant Foodworks: 3.13, and (10) Colgate Palmolive India: 2.98.

**Table 16.** The return and the risk of the FMCG sector stocks

| Stocks | Annual Return (%) | Annual Risk (%) |
|---|---|---|
| Hindustan Unilever | 32.35 | 23.20 |
| ITC | -2.71 | 27.57 |
| Nestle India | 32.27 | 24.90 |
| Tata Consumer Products | 65.03 | 35.26 |
| Britannia Industries | 27.76 | 26.46 |
| Dabur India | 18.25 | 24.87 |
| Godrej Consumer Products | 11.98 | 30.25 |
| Marico | 12.30 | 24.83 |
| Jubilant Foodworks | 60.71 | 38.90 |
| Colgate Palmolive India | 15.14 | 22.61 |

The annual return and the risk values for the stocks of the FMCG sector based on the historical prices from 1 January 2016 to 31 December 2020 are presented in Table 16. The highest return and the highest risk are found to be yielded by Tata Consumer Products and Jubilant Foodworks, respectively. On the other hand, ITC and Colgate Palmolive India exhibited the lowest return and the lowest risk, respectively.

Table 17. The portfolios of the FMCG sector stocks

| Stocks | Min Risk Portfolio | Opt. Risk Portfolio | Eigen Portfolio |
|---|---|---|---|
| Hindustan Unilever | 0.1525 | 0.0350 | 0.11 |
| ITC | 0.1961 | 0.0143 | 0.08 |
| Nestle India | 0.1750 | 0.1678 | 0.09 |
| Tata Consumer Products | 0.0254 | 0.2295 | 0.09 |
| Britannia Industries | 0.0654 | 0.1329 | 0.11 |
| Dabur India | 0.0295 | 0.0637 | 0.12 |
| Godrej Consumer Prod. | 0.0152 | 0.0067 | 0.11 |
| Marico | 0.1697 | 0.0724 | 0.11 |
| Jubilant Foodworks | 0.0083 | 0.2447 | 0.08 |
| Colgate Palmolive India | 0.1630 | 0.0331 | 0.10 |

Table 18. The return and the risk values of the FMCG sector portfolios

| Metric | Min Risk Portfolio | Opt. Risk Portfolio | Eigen Portfolio |
|---|---|---|---|
| Portfolio Return | 19.29 | 42.61 | 14.02 |
| Portfolio Risk | 16.41 | 20.92 | 13.23 |

Table 17 shows the allocation of weights to different stocks of the FMCG sector as per the computations of the three portfolio design approaches. The stocks which are assigned the highest weight by the minimum risk, the optimum risk, and the eigen portfolio are ITC, Jubilant Foodworks, and Dabur India, respectively.

Table 18 presents the risk and the return values associated with the portfolios of the FMCG durable stocks. As already mentioned, these values are computed using the historical prices of the stocks over the training period. The optimum risk portfolio is found to have produced the highest values for both return and risk for the stocks of the FMCG sector.

Table 19. The actual return of the optimum portfolio of the FMCG sector

| Stock | Date: January 1, 2021 | | | Date: July 1, 2021 | | Return |
|---|---|---|---|---|---|---|
|  | Actual Price | Amount Invested | No. of Stocks | Actual Price | Actual Value of Stocks |  |
| Hindustan Unilever | 2388 | 17120 | 7.17 | 2478 | 17765 |  |
| ITC | 214 | 7140 | 33.36 | 203 | 6773 |  |
| Nestle India | 18451 | 250 | 0.01 | 17646 | 239 |  |
| Tata Consumer Prod. | 602 | 3610 | 6.00 | 756 | 4533 |  |
| Britannia Industries | 3568 | 22340 | 6.26 | 3596 | 22515 |  |
| Dabur India | 534 | 4910 | 9.19 | 590 | 5425 | 9.53% |
| Godrej Consumer Prod. | 738 | 21390 | 28.98 | 888 | 25738 |  |
| Marico | 406 | 4680 | 11.53 | 535 | 6167 |  |
| Jubilant Foodworks | 2793 | 8740 | 3.13 | 3113 | 9741 |  |
| Colgate Palmolive India | 1578 | 9820 | 6.22 | 1708 | 10629 |  |
| Total |  | 100000 |  |  | 109525 |  |

The results for the optimum risk portfolio for the FMCG sector are presented in Table 19. Similar to the other sectors discussed previously, the number of shares for each stock purchased on 1 January 2021 is calculated using the actual share price of the stock and the amount of money invested in the stock. The total value of the stocks on 1 July 2021 is computed, and the overall return is derived. For the six-month period, i.e., from 1 January 2021 to 1 July 2021, the return for the FMCG sector optimum risk portfolio is found to be 9.53%. The efficient portfolio of the FMCG sector is shown in Figure 14.

**Table 20.** The actual return of the eigen portfolio of the FMCG sector

| Stock | Date: January 1, 2021 | | | Date: July 1, 2021 | | Return |
|---|---|---|---|---|---|---|
| | Price/Stock | Amount Invested | No. of Stocks | Price/Stock | Actual Value of Stocks | |
| Hindustan Unilever | 2388 | 11000 | 4.61 | 2478 | 11415 | |
| ITC | 214 | 8000 | 37.38 | 203 | 7589 | |
| Nestle India | 18451 | 9000 | 0.49 | 17646 | 8607 | |
| Tata Consumer Prod. | 602 | 9000 | 14.95 | 756 | 11302 | |
| Britannia Industries | 3568 | 11000 | 3.08 | 3596 | 11086 | 10.73% |
| Dabur India | 534 | 12000 | 22.47 | 590 | 13258 | |
| Godrej Consumer Prod. | 738 | 11000 | 14.91 | 888 | 13236 | |
| Marico | 406 | 11000 | 27.09 | 535 | 14495 | |
| Jubilant Foodworks | 2793 | 8000 | 2.86 | 3113 | 8917 | |
| Colgate Palmolive India | 1578 | 10000 | 6.34 | 1708 | 10824 | |
| Total | | 100000 | | | 110729 | |

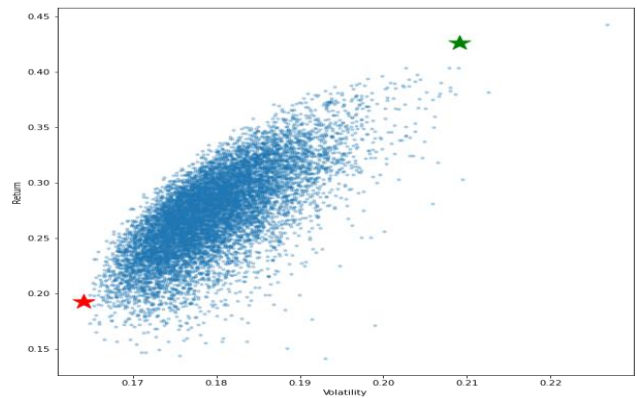

**Fig. 14.** The minimum risk portfolio (the red star) and the optimum risk portfolio (the green star) for the FMCG sector on historical stock prices from 1 January 2016 to 31 December 2020 (The risk is plotted along the *x*-axis and the return along the *y*-axis)

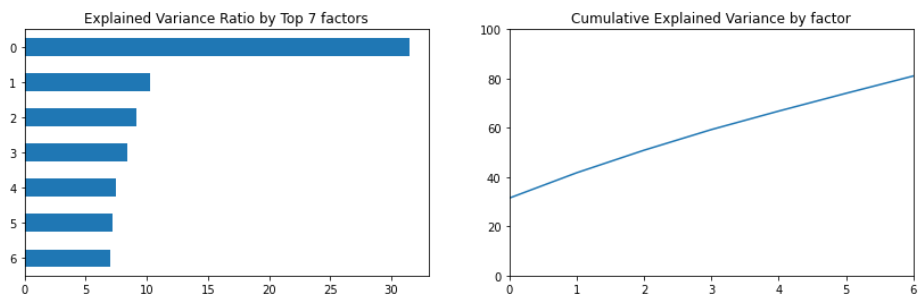

**Fig. 15.** The percentage of variance explained by seven components in the eigen portfolio and the cumulative explained variance by the same components of the eigen portfolio of the FMCG sector based on the historical prices of the stocks from 1 January 2016 to 31 December 2020.

Table 20 depicts the results of the eigen portfolio for the FMCG sector stocks. Seven components are used in building the eigen portfolio. The cumulative explanation of the variance by the seven components is 81.05%. Figure 15 depicts the variance and cumulative variance explained by the components. Figure 16 shows the correlation heatmap of the FMCG sector stocks, and Figure 17 depicts the weights assigned to the stocks of the FMCG sector by the five eigen portfolios, as discussed in Section 3.8. The eigen portfolio for the FMCG sector yielded a return of 10.73% over the six-month time horizon, which is marginally higher than the return of the optimum risk portfolio for the sector.

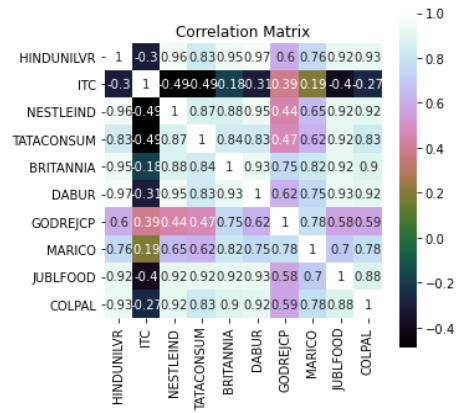

**Fig. 16.** The correlation heatmap for the FMCG sector stocks

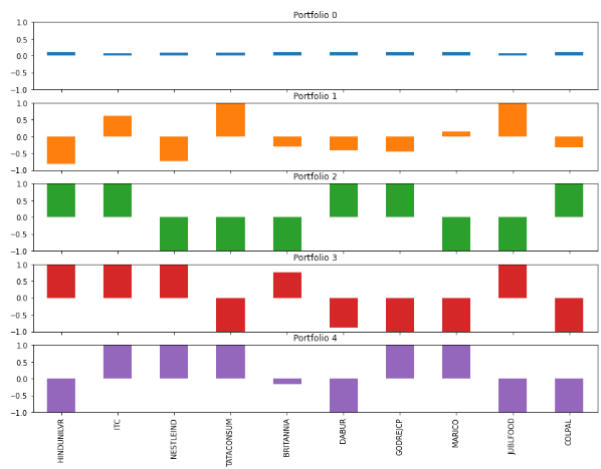

**Fig. 17.** Five eigen portfolios of the FMCG sector stocks with weights of each stock based on the first five principal components

### 4.5 Healthcare sector stocks

The top ten NSE-listed stocks of the healthcare sector and their respective weights in percent values in the computation of the overall index of the sector as per the report released by NSE on 30 June 2021 are as follows (NSE, 2021). (1) Sun Pharmaceuticals Industries: 14.35, (2) Dr. Reddy's Laboratories: 12.96, (3) Divi's Laboratories: 11.05, (4) Cipla: 9.72, (5) Apollo Hospitals Enterprise: 7.17, (6) Lupin: 5.44, (7) Aurobindo Pharma: 5.34, (8) Laurus Labs: 5.31, (9) Biocon: 3.63, (10) Cadila Healthcare: 3.25.

Table 21 presents the annual return and the risk values for the stocks constituting the healthcare sector based on their historical prices from 1 January 2016 to 31 December 2020. Laurus Labs and Aurobindo Pharma are found to have yielded the highest return and the highest risk, respectively. On the other hand, the lowest return and the lowest risk are exhibited by Lupin and Cipla, respectively.

Table 21. The return and the risk of the healthcare sector stocks

| Stocks | Annual Return (%) | Annual Risk (%) |
|---|---|---|
| Sun Pharma | 0.38 | 32.93 |
| Dr. Reddy's Labs | 19.25 | 28.93 |
| Divi's Labs | 51.37 | 36.07 |
| Cipla | 14.17 | 27.98 |
| Apollo Hospitals | 21.83 | 33.73 |
| Lupin | -6.65 | 31.45 |
| Aurobindo Pharma | 17.53 | 41.47 |
| Laurus Labs | 91.03 | 35.94 |
| Biocon | 34.08 | 35.05 |
| Cadila Healthcare | 15.85 | 33.91 |

The weights allocated to the stocks under three portfolio design approaches for the healthcare sector stocks are presented in Table 22. The stocks which are assigned the highest weight by the minimum risk, the optimum risk portfolios are Dr. Reddy's Labs and Divi's Labs, respectively. However, Sun Pharma, Apollo Hospitals, and Aurobindo Pharma received the highest weight in the eigen portfolio design.

The return and the risk values of the portfolios of the healthcare sector based on the historical prices of the stocks over the training period are exhibited in Table 23. The optimum risk portfolio is found to have produced the highest values for both return and risk for the stocks of the FMCG sector.

Table 22. The portfolios of the healthcare sector stocks

| Stocks | Min Risk Portfolio | Opt. Risk Portfolio | Eigen Portfolio |
|---|---|---|---|
| Sun Pharma | 0.0642 | 0.0207 | 0.1200 |
| Dr. Reddy's Labs | 0.1777 | 0.1409 | 0.1100 |
| Divi's Labs | 0.1016 | 0.3208 | 0.1000 |
| Cipla | 0.1610 | 0.0808 | 0.1100 |
| Apollo Hospitals | 0.1371 | 0.0100 | 0.0600 |
| Lupin | 0.1118 | 0.0225 | 0.1200 |
| Aurobindo Pharma | 0.0023 | 0.0726 | 0.1200 |
| Laurus Labs | 0.1066 | 0.2987 | 0.0700 |
| Biocon | 0.0586 | 0.0013 | 0.0900 |
| Cadila Healthcare | 0.0792 | 0.0315 | 0.1000 |

Table 23. The return and the risk values of the healthcare sector portfolios

| Metric | Min Risk Portfolio | Opt. Risk Portfolio | Eigen Portfolio |
|---|---|---|---|
| Portfolio Return | 26.19 | 49.43 | 25 |
| Portfolio Risk | 20.39 | 24.06 | 17 |

The results of the optimum risk portfolio for the healthcare sector are presented in Table 24. For the six-month period, i.e., from 1 January 2021 to 1 July 2021, the optimum risk portfolio of the healthcare sector yielded a return of 36.28%. The efficient portfolio of the healthcare sector is shown in Figure 18.

**Table 24.** The actual return of the optimum portfolio of the healthcare sector

| Stock | Date: January 1, 2021 | | | Date: July 1, 2021 | | Return |
|---|---|---|---|---|---|---|
| | Price/Stock | Amount Invested | No. of Stocks | Price/Stock | Actual Value of Stocks | |
| Sun Pharma | 596 | 2070 | 3.47 | 684 | 2376 | |
| Dr. Reddy's Labs | 5241 | 14090 | 2.69 | 5559 | 14945 | |
| Divi's Labs | 3849 | 32080 | 8.33 | 4436 | 36972 | |
| Cipla | 827 | 8080 | 9.77 | 978 | 9555 | |
| Apollo Hospitals | 2415 | 1000 | 0.41 | 3678 | 1523 | 36.28% |
| Lupin | 1001 | 2250 | 2.25 | 1146 | 2576 | |
| Aurobindo Pharma | 928 | 7260 | 7.82 | 968 | 7573 | |
| Laurus Labs | 353 | 29870 | 84.62 | 667 | 56440 | |
| Biocon | 466 | 150 | 0.28 | 406 | 113 | |
| Cadila Healthcare | 478 | 3150 | 6.59 | 639 | 4211 | |
| Total | | 100000 | | | 136284 | |

**Table 25.** The actual return of the eigen portfolio of the healthcare sector

| Stock | Date: January 1, 2021 | | | Date: July 1, 2021 | | Return |
|---|---|---|---|---|---|---|
| | Price/Stock | Amount Invested | No. of Stocks | Price/Stock | Actual Value of Stocks | |
| Sun Pharma | 596 | 12000 | 20.13 | 684 | 13772 | |
| Dr. Reddy's Labs | 5241 | 11000 | 2.10 | 5559 | 11667 | |
| Divi's Labs | 3849 | 10000 | 2.60 | 4436 | 11525 | |
| Cipla | 827 | 11000 | 13.30 | 978 | 13008 | |
| Apollo Hospitals | 2415 | 6000 | 2.48 | 3678 | 9138 | 19.80% |
| Lupin | 1001 | 12000 | 11.99 | 1146 | 13738 | |
| Aurobindo Pharma | 928 | 12000 | 12.93 | 968 | 12517 | |
| Laurus Labs | 353 | 7000 | 19.83 | 667 | 13227 | |
| Biocon | 466 | 9000 | 19.31 | 406 | 7841 | |
| Cadila Healthcare | 478 | 10000 | 20.92 | 639 | 13368 | |
| Total | | 100000 | | | 119801 | |

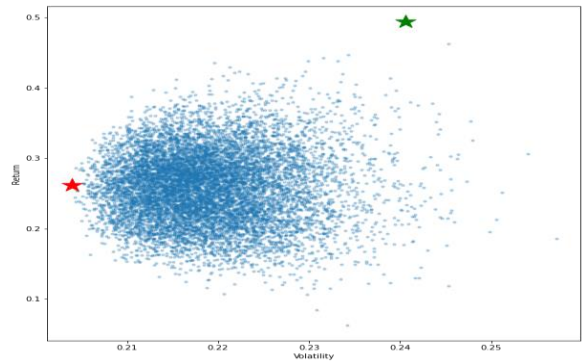

**Fig. 18.** The minimum risk portfolio (the red star) and the optimum risk portfolio (the green star) for the healthcare sector on historical stock prices from 1 January 2016 to 31 December 2020 (The risk is plotted along the *x*-axis and the return along the *y*-axis)

Table 25 shows the results of the eigen portfolio for the healthcare sector. The cumulative explanation of the variance by the seven components in the eigen portfolio is 83.98%. Figure 19 depicts the variance and cumulative variance explained by the seven components. Figure 20 shows the correlation heatmap of the healthcare sector stocks, and Figure 21 depicts the weights assigned to the stocks of the healthcare sector by the five eigen portfolios, as discussed in Section 3.8. The eigen portfolio for the healthcare sector yields a return of 19.80% for the six-month period (i.e., 1 January 2016 to 31 December 2020). The return of the eigen portfolio is substantially lower than the corresponding figure of the optimum risk portfolio for the sector.

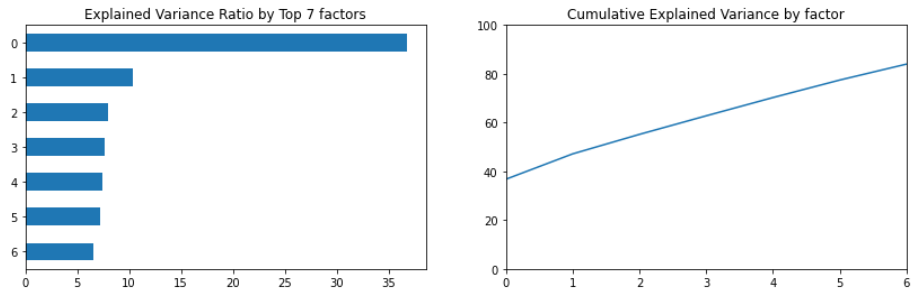

**Fig. 19.** The percentage of variance explained by seven components in the eigen portfolio and the cumulative explained variance by the same components of the eigen portfolio of the healthcare sector based on the historical prices of the stocks from 1 January 2016 to 31 December 2020.

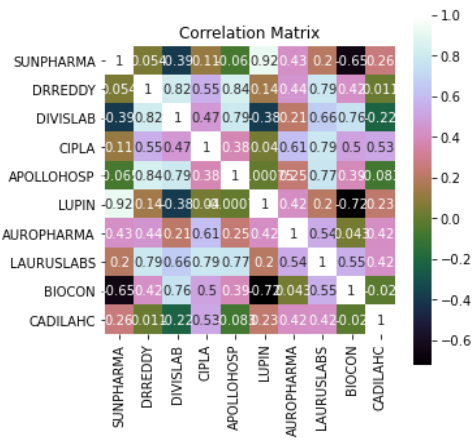

**Fig. 20.** The correlation heatmap for the healthcare sector stocks

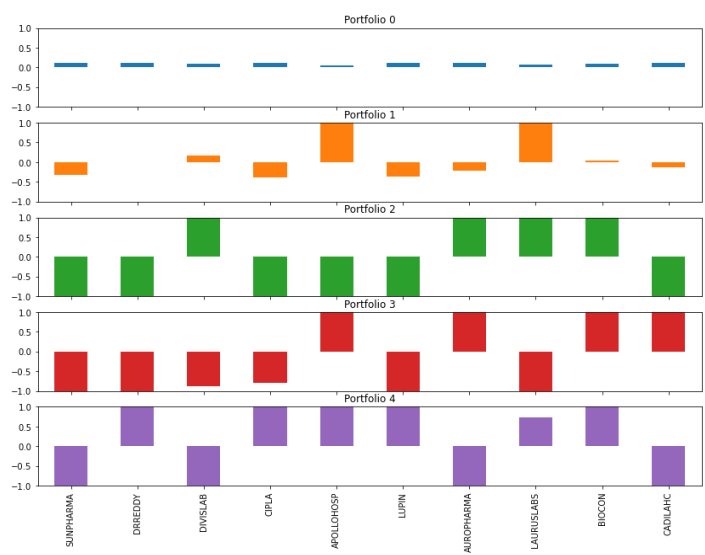

**Fig. 21.** Five eigen portfolios of the healthcare sector stocks with weights of each stock based on the first five principal components

## 4.6 IT sector stocks

The top ten NSE-listed stocks of the IT sector and their respective weights in percent values in the computation of the overall index of the sector as per the report released by NSE on 30 June 2021 are as follows (NSE, 2021). (1) Infosys: 26.63, (2) Tata Consultancy Services (TCS): 26.25, (3) Tech Mahindra: 9.10, (4) Wipro: 9.02, (5) HCL Technologies: 8.94, (6) Larsen & Toubro Infotech (L & T Infotech): 5.30, (7) MphasiS: 5.03, (8) MindTree: 4.78, (9) Coforge: 2.52, and (10) Oracle Financial Services Software (OFSS): 2.43.

The annual return and risk values for the IT sector stocks based on their historical prices from 1 January 2016 to 31 December 2020 are exhibited in Table 26. Coforge is found to have yielded the highest values for both return and risk. On the other hand, the lowest return and the lowest risk are exhibited by OFSS and Wipro, respectively.

**Table 26.** The return and the risk of the IT sector stocks

| Stocks | Annual Return (%) | Annual Risk (%) |
|---|---|---|
| Infosys | 27.77 | 28.54 |
| TCS | 25.78 | 25.81 |
| Tech Mahindra | 20.24 | 31.01 |
| Wipro | 23.30 | 25.47 |
| HCL Technologies | 24.81 | 27.54 |
| L & T Infotech | 57.38 | 33.15 |
| MphasiS | 32.93 | 33.65 |
| MindTree | 38.71 | 35.94 |
| Coforge | 59.47 | 43.22 |
| OFSS | 3.30 | 27.45 |

**Table 27.** The portfolios of the IT sector stocks

| Stocks | Min Risk Portfolio | Opt. Risk Portfolio | Eigen Portfolio |
|---|---|---|---|
| Infosys | 0.0426 | 0.0546 | 0.13 |
| TCS | 0.2165 | 0.2051 | 0.12 |
| Tech Mahindra | 0.0479 | 0.0013 | 0.12 |
| Wipro | 0.1327 | 0.0419 | 0.11 |
| HCL Technologies | 0.1289 | 0.0311 | 0.13 |
| L & T Infotech | 0.0659 | 0.2300 | 0.07 |
| MphasiS | 0.1256 | 0.2089 | 0.06 |
| MindTree | 0.0225 | 0.0802 | 0.10 |
| Coforge | 0.0028 | 0.1381 | 0.10 |
| OFSS | 0.2145 | 0.0086 | 0.06 |

**Table 28.** The return and the risk values of the IT sector portfolios

| Metric | Min Risk Portfolio | Opt. Risk Portfolio | Eigen Portfolio |
|---|---|---|---|
| Portfolio Return | 23.69 | 40.00 | 19.38 |
| Portfolio Risk | 18.38 | 21.60 | 13.06 |

Table 27 shows the allocation of weights to the stocks of the IT sector under three portfolio design approaches. The stocks which are assigned the highest weight by the minimum risk and the optimum risk portfolios are TCS and Larsen & Toubro Infotech, respectively. However, Infosys and HCL Technologies are allocated the highest weight under the eigen portfolio.

Table 28 presents the return and the risk values of the portfolios of the IT sector using the historical prices of the stocks over the training period. The optimum risk portfolio is found to have produced the highest values for both return and risk for the IT sector stocks.

**Table 29.** The actual return of the optimum portfolio of the IT sector

| Stock | Date: January 1, 2021 | | | Date: July 1, 2021 | | Return |
|---|---|---|---|---|---|---|
| | Price/Stock | Amount Invested | No. of Stocks | Price/Stock | Actual Value of Stocks | |
| Infosys | 1260 | 5460 | 4.33 | 1560 | 6760 | 28.71% |
| TCS | 2928 | 20530 | 7.00 | 3342 | 23410 | |
| Tech Mahindra | 978 | 130 | 0.13 | 1085 | 144 | |
| Wipro | 388 | 4190 | 10.80 | 539 | 5821 | |
| HCL Technologies | 951 | 3110 | 3.27 | 986 | 3224 | |
| L & T Infotech | 3699 | 23000 | 6.22 | 4017 | 24977 | |
| MphasiS | 1530 | 20890 | 13.65 | 2172 | 29656 | |
| MindTree | 1659 | 8020 | 4.83 | 2581 | 12477 | |
| Coforge | 2722 | 13810 | 5.07 | 4194 | 21278 | |
| OFSS | 3243 | 860 | 0.27 | 3620 | 960 | |
| Total | | 100000 | | | 128707 | |

Table 29 shows the results for the optimum risk portfolio of the IT sector stocks. The return of the optimum portfolio for the IT sector stocks over the six-month period, i.e., from 1 January 2021 to 1 July 2021, is found to be 28.71%. The efficient portfolio of the IT sector is shown in Figure 22.

**Table 30.** The actual return of the eigen portfolio of the IT sector

| Stock | Date: January 1, 2021 | | | Date: July 1, 2021 | | Return |
|---|---|---|---|---|---|---|
| | Price/Stock | Amount Invested | No. of Stocks | Price/Stock | Actual Value of Stocks | |
| Infosys | 1260 | 13000 | 10.32 | 1560 | 16095 | 25.65% |
| TCS | 2928 | 12000 | 4.10 | 3342 | 13697 | |
| Tech Mahindra | 978 | 12000 | 12.27 | 1085 | 13313 | |
| Wipro | 388 | 11000 | 28.35 | 539 | 15281 | |
| HCL Technologies | 951 | 13000 | 13.67 | 986 | 13478 | |
| L & T Infotech | 3699 | 7000 | 1.89 | 4017 | 7602 | |
| MphasiS | 1530 | 6000 | 3.92 | 2172 | 8518 | |
| MindTree | 1659 | 10000 | 6.03 | 2581 | 15558 | |
| Coforge | 2722 | 10000 | 3.67 | 4194 | 15408 | |
| OFSS | 3243 | 6000 | 1.85 | 3620 | 6698 | |
| Total | | 100000 | | | 125648 | |

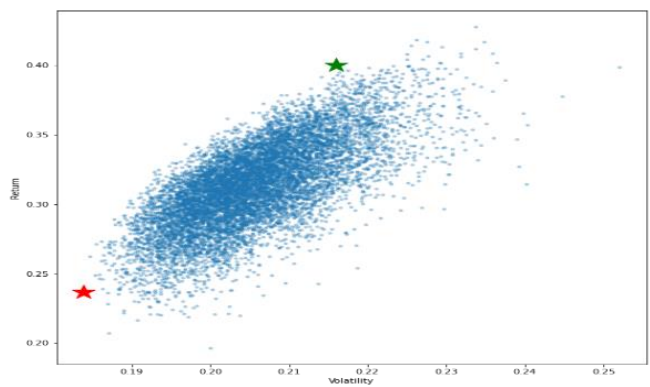

**Fig. 22.** The minimum risk portfolio (the red star) and the optimum risk portfolio (the green star) for the IT sector on historical stock prices from 1 January 2016 to 31 December 2020 (The risk is plotted along the *x*-axis and the return along the *y*-axis)

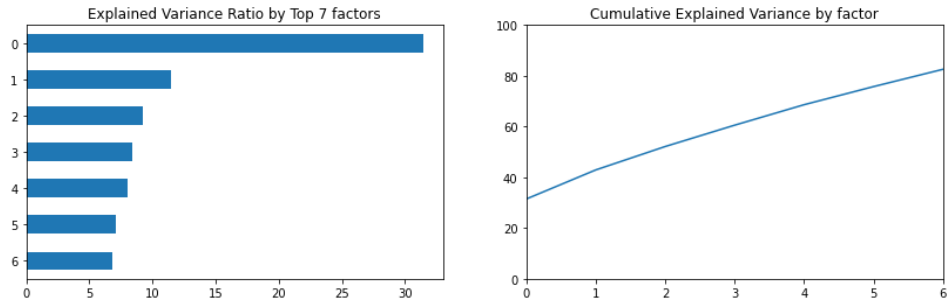

**Fig. 23.** The percentage of variance explained by seven components in the eigen portfolio and the cumulative explained variance by the same components of the eigen portfolio of the IT sector based on the historical prices of the stocks from 1 January 2016 to 31 December 2020.

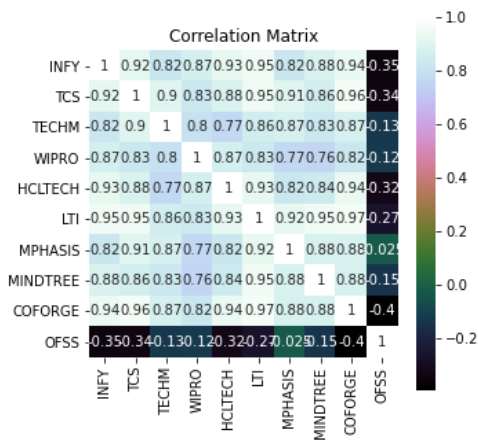

**Fig. 24.** The correlation heatmap for the IT sector stocks

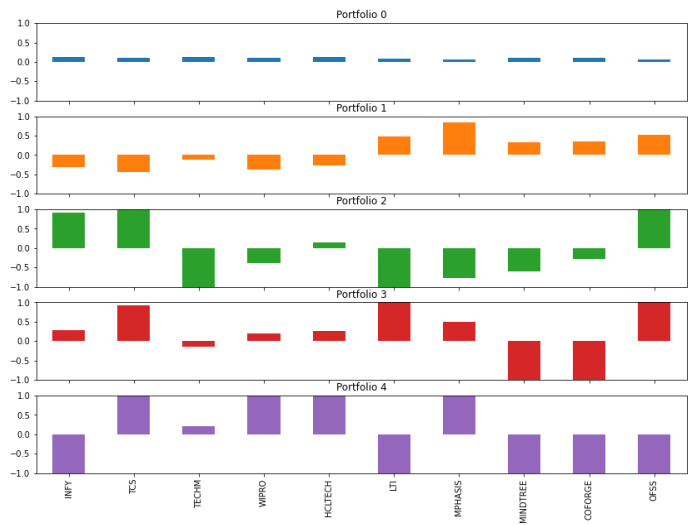

**Fig. 25.** Five eigen portfolios of the IT sector stocks with weights of each stock based on the first five principal components

Table 30 presents the results of the eigen portfolio for the IT sector. The cumulative explanation of the variance by the seven components in the eigen portfolio is 82.58%. Figure 23 depicts the variance and cumulative variance explained by the seven components. Figure 24 shows the correlation heatmap of the IT sector stocks, and Figure 25 depicts the weights assigned to the stocks of the IT sector by the five eigen portfolios, as discussed in Section 3.8. The eigen portfolio for the IT sector over the six-month period yielded a return of 25.65%. The return of the eigen portfolio is lower than the corresponding figure of the optimum risk portfolio.

### 4.7 Metal sector stocks

The top ten NSE-listed stocks of the metal sector and their respective weights in percent values in the computation of the overall index of the sector as per the report released by NSE on 30 June 2021 are as follows (NSE, 2021). (1) Tata Steel: 22.64, (2) JSW Steel: 16.42, (3) Hindalco Industries: 13.49, (4) Adani Enterprises: 10.29, (5) Vedanta: 8.25, (6) Coal India: 7.63, (7) Steel Authority of India (SAIL): 4.49, (8) NMDC: 4.29, (9) Jindal Steel & Power: 4.03, and (10) APL Apollo Tubes: 2.63.

Table 31 exhibits the annual return and risk values for the metal sector stocks based on their historical prices from 1 January 2016 to 31 December 2020. Coal India is found to have yielded the lowest values for both return and risk. However, the highest return and the highest risk are exhibited by Adani Enterprises and Jindal Steel & Power, respectively.

**Table 31.** The return and the risk of the metal sector stocks

| Stocks | Annual Return (%) | Annual Risk (%) |
|---|---|---|
| Tata Steel | 22.03 | 38.05 |
| JSW Steel | 27.62 | 36.24 |
| Hindalco Industries | 16.13 | 41.66 |
| Adani Enterprises | 89.18 | 51.25 |
| Vedanta | -1.10 | 47.41 |
| Coal India | -17.21 | 29.39 |
| SAIL | 21.70 | 46.27 |
| NMDC | 1.24 | 39.23 |
| Jindal Steel & Power | 59.25 | 57.06 |
| APL Apollo Tubes | 64.77 | 37.88 |

The allocation of weights to different stocks for the metal sectors stocks under three portfolio design approaches is presented in Table 32. The stocks which are assigned the highest weight by the minimum risk and the optimum risk portfolios are Coal India and APL Apollo Tubes, respectively. The eigen portfolio has assigned the highest weight to Tata Steel and SAIL, however.

**Table 32.** The portfolios of the metal sector stocks

| Stocks | Min Risk Portfolio | Opt. Risk Portfolio | Eigen Portfolio |
|---|---|---|---|
| Tata Steel | 0.1228 | 0.1168 | 0.12 |
| JSW Steel | 0.2561 | 0.0597 | 0.11 |
| Hindalco Industries | 0.0531 | 0.0472 | 0.11 |
| Adani Enterprises | 0.0568 | 0.2236 | 0.08 |
| Vedanta | 0.0070 | 0.0151 | 0.11 |
| Coal India | 0.2700 | 0.0523 | 0.07 |
| SAIL | 0.0134 | 0.0061 | 0.12 |
| NMDC | 0.1014 | 0.0336 | 0.10 |
| Jindal Steel & Power | 0.0019 | 0.1828 | 0.11 |
| APL Apollo Tubes | 0.1174 | 0.2629 | 0.07 |

**Table 33.** The return and the risk values of the metal sector portfolios

| Metric | Min Risk Portfolio | Opt. Risk Portfolio | Eigen Portfolio |
|---|---|---|---|
| Portfolio Return | 19.18 | 52.04 | 55.35 |
| Portfolio Risk | 25.98 | 31.76 | 46.64 |

Table 33 shows the return and the risk values of the portfolios of the metal sector using the historical prices of the stocks over the training period. The eigen portfolio is found to have yielded the highest values for both return and risk for the stocks of the metal sector.

**Table 34.** The actual return of the optimum portfolio of the metal sector

| Stock | Date: January 1, 2021 | | | Date: July 1, 2021 | | Return |
|---|---|---|---|---|---|---|
| | Price/Stock | Amount Invested | No. of Stocks | Price/Stock | Actual Value of Stocks | |
| Tata Steel | 643 | 11680 | 18.16 | 1164 | 21144 | |
| JSW Steel | 390 | 5970 | 15.31 | 681 | 10425 | |
| Hindalco Industries | 238 | 4720 | 19.83 | 379 | 7516 | |
| Adani Enterprises | 491 | 22360 | 45.54 | 1490 | 67854 | |
| Vedanta | 160 | 1500 | 9.44 | 263 | 2482 | 98.41% |
| Coal India | 135 | 5230 | 38.74 | 146 | 5656 | |
| SAIL | 75 | 610 | 8.13 | 127 | 1033 | |
| NMDC | 116 | 3360 | 28.97 | 184 | 5330 | |
| Jindal Steel & Power | 270 | 18280 | 67.70 | 395 | 26743 | |
| APL Apollo Tubes | 860 | 26290 | 30.57 | 1643 | 50226 | |
| Total | | 100000 | | | 198409 | |

**Table 35.** The actual return of the eigen portfolio of the metal sector

| Stock | Date: January 1, 2021 | | | Date: July 1, 2021 | | Return |
|---|---|---|---|---|---|---|
| | Price/Stock | Amount Invested | No. of Stocks | Price/Stock | Actual Value of Stocks | |
| Tata Steel | 643 | 12000 | 18.66 | 1164 | 21723 | |
| JSW Steel | 390 | 11000 | 28.21 | 681 | 19208 | |
| Hindalco Industries | 238 | 11000 | 46.22 | 379 | 17517 | |
| Adani Enterprises | 491 | 8000 | 16.29 | 1490 | 24277 | |
| Vedanta | 160 | 11000 | 68.75 | 263 | 18081 | 74.02% |
| Coal India | 135 | 7000 | 51.85 | 146 | 7570 | |
| Steel Auth. of India | 75 | 12000 | 160.00 | 127 | 20320 | |
| NMDC | 116 | 10000 | 86.21 | 184 | 15862 | |
| Jindal Steel & Power | 270 | 11000 | 40.74 | 395 | 16093 | |
| APL Apollo Tubes | 860 | 7000 | 8.14 | 1643 | 13373 | |
| Total | | 100000 | | | 174024 | |

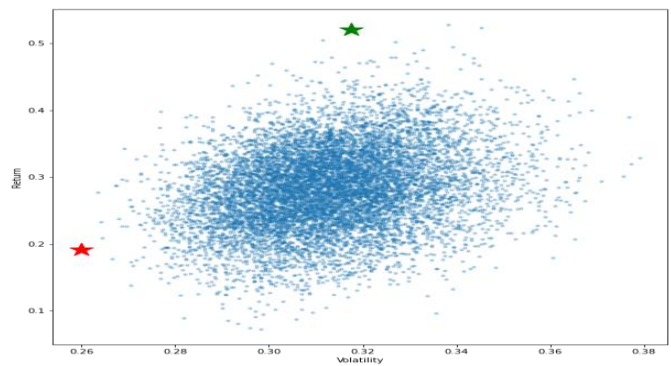

**Fig. 26.** The minimum risk portfolio (the red star) and the optimum risk portfolio (the green star) for the metal sector on historical stock prices from 1 January 2016 to 31 December 2020 (The risk is plotted along the *x*-axis and the return along the *y*-axis)

The results for the optimum risk portfolio of the metal sector stocks are depicted in Table 34. The return of the optimum portfolio for the metal sector stocks over the six-month period, i.e., from 1 January 2021 to 1 July 2021, is found to be 98.41%. The efficient portfolio of the metal sector is shown in Figure 26.

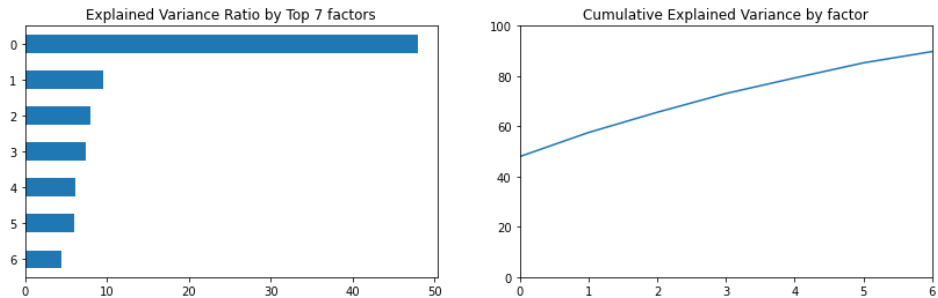

**Fig. 27.** The percentage of variance explained by seven components in the eigen portfolio and the cumulative explained variance by the same components of the eigen portfolio of the metal sector based on the historical prices of the stocks from 1 January 2016 to 31 December 2020.

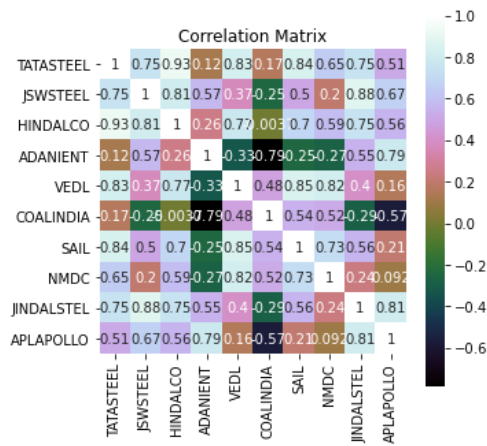

**Fig. 28.** The correlation heatmap for the metal sector stocks

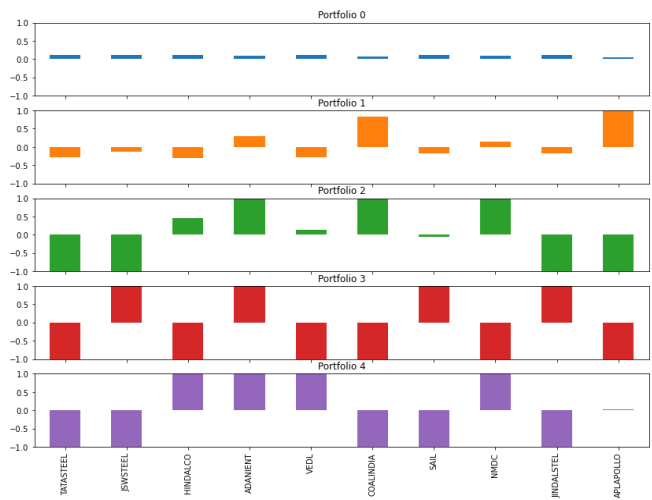

**Fig. 29.** Five eigen portfolios of the metal sector stocks with weights of each stock based on the first five principal components

Table 35 presents the results of the eigen portfolio for the metal sector. The cumulative explanation of the variance by the seven components in the eigen portfolio is found to be 89.76%. Figure 27 illustrates the variance and cumulative variance explained by the seven components. Figure 28 shows the correlation heatmap of the metal sector stocks, and Figure 29 depicts the weights assigned to the stocks of the metal sector by the five eigen portfolios, as discussed in Section 3.8. The eigen portfolio for the metal sector has produced an overall return of 74.02% over the six-month period. The return of the eigen portfolio is lower than the corresponding figure of the optimum risk portfolio.

### 4.8 Summary of the results

The results of the seven sectors are summarized in Table 36. It is observed that the highest return is yielded by the optimum risk portfolio for the metal sector. The FMCG sector under the optimum risk portfolio produces the lowest return. Another interesting observation is that among the seven sectors, the optimum risk portfolio has yielded a higher return for four of them, while the eigen portfolio has produced a higher return for the other three sectors. The sectors for which the return yielded by the optimum risk portfolio is higher are consumer durable, healthcare, IT, and metal. The sectors for which the eigen portfolio yields higher returns are auto, banking, and FMCG. Hence, the performances of the two portfolio design approaches have been almost similar, and we have no reason to prefer one approach over the other. However, the performance of the optimum risk portfolio is found to be marginally better for the seven sectors analyzed in this work.

**Table 36.** The return in six months on two portfolios for different sectors

| Sector | Return of opt portfolio (%) | Return of eigen portfolio (%) |
|---|---|---|
| Auto | 18.17 | 21.52 |
| Banking | 8.66 | 18.54 |
| Consumer Durable | 27.77 | 18.49 |
| FMCG | 9.53 | 10.73 |
| Healthcare | 36.28 | 19.80 |
| IT | 28.71 | 25.65 |
| Metal | 98.41 | 74.02 |

### 5. Conclusion

Designing a portfolio for optimizing the return and risk is a very challenging task. In this paper, we have presented three different approaches to portfolio design, e.g., the minimum risk portfolio, the optimum risk portfolio, and the eigen portfolio. We have chosen seven important sectors in the Indian stock market, and identified the top ten stocks from each sector based on their listing in the NSE of India. Using the historical stock prices for each stock from 1 January 2016 to 31 December 2020, three portfolios are designed for each sector. Based on the training data, several important metrics of each portfolio are computed such as, annual return, annual risk, weight assigned to the constituent stocks in a portfolio, the correlation heatmaps among the stocks, and the principal components of the eigen portfolios. After evaluating the portfolios on their training data, we deployed the optimum risk portfolios and the eigen risk portfolios over a period of six months and computed their return values. In other words, the optimum and the eigen portfolios for each of the seven sectors designed on 1 January 2021, are evaluated for their return values on 1 July 2021. It is observed that the return values of the optimum risk portfolio for the consumer durable, healthcare, IT, and metal sectors are higher than those of the eigen portfolios. On the other hand, the eigen portfolio has yielded higher returns for auto, banking, and the FMCG sectors. While the performances of the portfolio design approaches are found to be similar, it is observed the return values of the optimum risk portfolio is marginally higher than its eigen counterparts. For the optimum risk portfolio, the highest return is found for the metal sector, while the return for the banking sector is the lowest. On the other hand, the eigen portfolio has yielded the highest return for the metal sector, and the lowest return for the FMCG sector.

As a future plan of research, we intend to use reinforcement learning algorithms e.g., Q-learning, and Deep Q-learning to introduce more robustness in the portfolios. Including additional sectors in the analysis framework for designing more comprehensive portfolios is also another direction of future research.